\documentstyle[preprint,aps,epsf]{revtex}
\tightenlines
\begin{document}
\title{Scalar Quarkonium Masses and Mixing with the Lightest Scalar Glueball}
\author{W.\ Lee \cite{LANL}
and D.\ Weingarten}
\address{ IBM Research, P.O.~Box 218,
Yorktown Heights, NY 10598}
\maketitle

\begin{abstract}

We evaluate the continuum limit of the valence (quenched) approximation
to the mass of the lightest scalar quarkonium state, for a range of
different quark masses, and to the mixing energy between these states
and the lightest scalar glueball. Our results support the interpretation
of $f_0(1710)$ as composed mainly of the lightest scalar glueball.
 
\end{abstract}
\narrowtext

Evidence that $f_0(1710)$ is composed mainly of the lightest scalar
glueball is now given by two different sets of numerical determinations
of QCD predictions using the theory's lattice formulation in the valence
(quenched) approximation.  A calculation on GF11~\cite{Sexton95} of the
width for the lightest scalar glueball to decay to all possible
pseudoscalar pairs, on a $16^3 \times 24$ lattice with $\beta$ of 5.7,
corresponding to a lattice spacing $a$ of 0.140(4) fm, gives 108(29)
MeV.  This number combined with any reasonable guesses for the effect of
finite lattice spacing, finite lattice volume, and the remaining width
to multibody states yields a total width small enough for the lightest
scalar glueball to be seen easily in experiment.  For the infinite
volume continuum limit of the lightest scalar glueball mass, a
reanalysis~\cite{latestglue} of a calculation on GF11~\cite{Vaccarino},
using 25000 to 30000 gauge configurations, gives 1648(58) MeV.  An
independent calculation by the UKQCD-Wuppertal~\cite{Livertal}
collaboration, using 1000 to 3000 gauge configurations, when
extrapolated to the continuum limit according to
Ref.~\cite{latestglue,Weingarten94} yields 1568(89) MeV.  A more recent
calculation using an improved action~\cite{Morningstar} gives 1630(100)
MeV.  The three results combined become 1632(49) MeV. A phenomenological
model of the glueball spectrum which supports this prediction is
discussed in Ref.~\cite{Brisudova}.

Among established resonances with the quantum numbers to be a scalar
glueball, all are clearly inconsistent with the mass calculations except
$f_0(1710)$ and $f_0(1500)$.  Between these two, $f_0(1710)$ is favored
by the mass result with largest statistics, by the combined result, and
by the expectation~\cite{Weingarten97} that the valence approximation
will lead to an underestimate of the scalar glueball's mass.
Refs.~\cite{Sexton95,Weingarten97} interpret $f_0(1500)$ as dominantly
composed of strange-antistrange, $s\overline{s}$, scalar quarkonium.  A
possible objection to this interpretation, however, is that $f_0(1500)$
apparently does not decay mainly to states containing an $s$ and an
$\overline{s}$ quark~\cite{Amsler1}. In part for this reason,
Ref.~\cite{Amsler2} interprets $f_0(1500)$ as composed mainly of the
lightest scalar glueball and $f_0(1710)$ as largely $s\overline{s}$
scalar quarkonium.  A second objection is that while the Hamiltonian of
full QCD couples quarkonium and glueballs, so that physical states
should be linear combinations of both, mixing is not treated
quantitatively in Ref.~\cite{Sexton95}. In the extreme, mixing could
lead to $f_0(1710)$ and $f_0(1500)$ each half glueball and half
quarkonium.

Using the valence approximation for a fixed lattice period $L$ of about
1.6 fm and a range of different values of quark mass, we have now
calculated the continuum limit of the mass of the lightest scalar
$q\overline{q}$ states and the continuum limit of the mixing energy
between these states and the lightest scalar glueball.  The continuum
values are obtained by extrapolation of results obtained from four
different values of lattice spacing. For the two largest lattice
spacings we have also done calculations on lattices with $L$ of about
2.3 fm. Preliminary versions of this work are reported in
Refs.~\cite{Weingarten97,Weingarten98,Lee97}.

Our results provide answers to the objections to the interpretation of
$f_0(1710)$ as largely the lightest scalar glueball.  For the valence
approximation to the infinite volume continuum limit of the
$s\overline{s}$ scalar mass we find a value significantly below the
valence approximation scalar glueball mass. This prediction rules out,
in our opinion, the possibility of identifying~\cite{Amsler2}
$f_0(1500)$ as primarily a glueball and $f_0(1710)$ as primarily
$s\overline{s}$ quarkonium. Our calculation of glueball-quarkonium
mixing energy, combined with the simplification of considering mixing
only among the lightest discrete isosinglet scalar states, then yields a
mixed $f_0(1710)$ which is 73.8(9.5)\% glueball and a mixed $f_0(1500)$
which is 98.4(1.4)\% quarkonium, mainly $s\overline{s}$. The glueball
amplitude which leaks from $f_0(1710)$ goes almost entirely to the state
$f_0(1390)$, which remains mainly $(u\overline{u} +
d\overline{d})/\sqrt{2}$. For $(u\overline{u} + d\overline{d})/\sqrt{2}$
in the rest of this article we will use the abbreviation
$n\overline{n}$, normal-antinormal. We find, in addition, that
$f_0(1500)$ acquires an $n\overline{n}$ amplitude with sign opposite to
its $s\overline{s}$ component suppressing, by interferences, the state's
decay to $K\overline{K}$ final states.  Assuming SU(3) flavor symmetry
before mixing for the decay couplings of scalar quarkonium to pairs of
pseudoscalars, the $K\overline{K}$ decay rate of $f_0(1500)$ is
suppressed by a factor of 0.39(16) in comparison to the rate of an
unmixed $s\overline{s}$ scalar.  This suppression is consistent, within
uncertainties, with the experimentally observed suppression.

Our calculations, using Wilson fermions and the plaquette action, were
done with ensembles of 2749 configurations on a lattice $12^2 \times 10
\times 24$ with $\beta$ of 5.70, 1972 configurations on $16^3 \times 24$ with
$\beta$ of 5.70, 2328 configurations on $ 16^2 \times 14 \times 20$ with
$\beta$ of 5.93, 1733 configurations on $24^4$ at $\beta$ of 5.93, 1000
configurations on $24^2 \times 20
\times 32$ with $\beta$ of 6.17, and 1003 configurations on $32^2 \times
28 \times 40$ with $\beta$ of 6.40. For $\beta$ of 5.70, 5.93, 6.17 and
6.40 the corresponding values of lattice spacing $a$ are, respectively,
0.140(4) fm, 0.0961(25) fm, 0.0694(18) fm, and 0.0519(14)
fm.  The smaller lattices with $\beta$ of 5.70 and 5.93, and the
lattices with $\beta$ of 6.17 and 6.40 have periods in the two (or
three) equal space directions of 1.68(5) fm, 1.54(4) fm, 1.74(5)
fm, 1.66(5) fm, respectively, and thereby permit extrapolations to
zero lattice spacing with nearly constant physical volume.

These values of lattice spacing, and conversions from lattice to physical
units in the remainder of this article, are
determined~\cite{latestglue,Vaccarino} from the exact solution to the
two-loop zero-flavor Callan-Symanzik equation for
$\Lambda^{(0)}_{\overline{MS}} a$ with $\Lambda^{(0)}_{\overline{MS}}$
of 234.9(6.2) MeV determined from the continuum limit of
$(\Lambda^{(0)}_{\overline{MS}} a)/(m_{\rho} a)$ in Ref.~\cite{Butler}.
For $\beta$ from 5.70 to 6.17, the ratio $(\Lambda^{(0)}_{\overline{MS}}
a)/(m_{\rho} a)$ in Ref.~\cite{Butler} was found to be constant within
statistical errors, thus our results are, within errors, almost
certainly the same as those we would have obtained by converting to
physical units using values of $m_{\rho} a$.  We chose to convert using
$\Lambda^{(0)}_{\overline{MS}} a$, however, since Ref.~\cite{Butler} did
not find $m_{\rho} a$ at $\beta$ of 6.40, which would be needed
for our present calculations.

For each ensemble of gauge fields, we evaluated correlation functions
using random sources built from quark and antiquark fields following
Ref~\cite{Lee97}.  Averaged over random sources, these correlation
functions become
\begin{eqnarray}
\label{defC}
C_{ff}(t) & = & \sum_{\vec{x}} < f(\vec{x},t) f(0,0)>, \nonumber \\
C_{gs}(t) & = & \sum_{\vec{x}} < g(\vec{x},t) s(0,0)>. 
\end{eqnarray}
Here $f$ is either $p$, $s$ or $g$. The quantities
$p(\vec{x},t)$ and $s(\vec{x},t)$ are, respectively, the smeared
pseudoscalar and scalar operators of Ref.~\cite{Lee97} built from a
quark and an antiquark field and $g(\vec{x},t)$ is the smeared scalar
glueball operator of Ref.~\cite{Vaccarino}. 

Fitting the the large-$t$ behavior of the diagonal correlators to the
asymptotic form
\begin{eqnarray}
\label{diagZm}
C_{ff}(t) & \rightarrow & Z_f exp(-m_f a t),
\end{eqnarray}
we obtained the masses, in lattice units, $m_p a$, $m_s a$, 
and $m_g a$ and field strength renormalization constants $Z_p$, $Z_s$ and
$Z_g$. From the large-$t$ behavior of the off-diagonal correlator, for
$m_s$ close to $m_g$,
\begin{eqnarray}
\label{offdiag}
C_{gs}(t) \rightarrow 
\sqrt{Z_g Z_{s}} E a   
\sum_{t'} exp( -m_g a |t - t'| - m_s a | t'|) 
\end{eqnarray}
we then found the glueball-quarkonium mixing energy, in lattice units, $E a$.
Eqs.~(\ref{diagZm}) and (\ref{offdiag}) have been simplified by omitting
terms arising from propagation around the lattice's periodic time
boundary.

\begin{figure}
\epsfxsize=\textwidth
\epsfbox{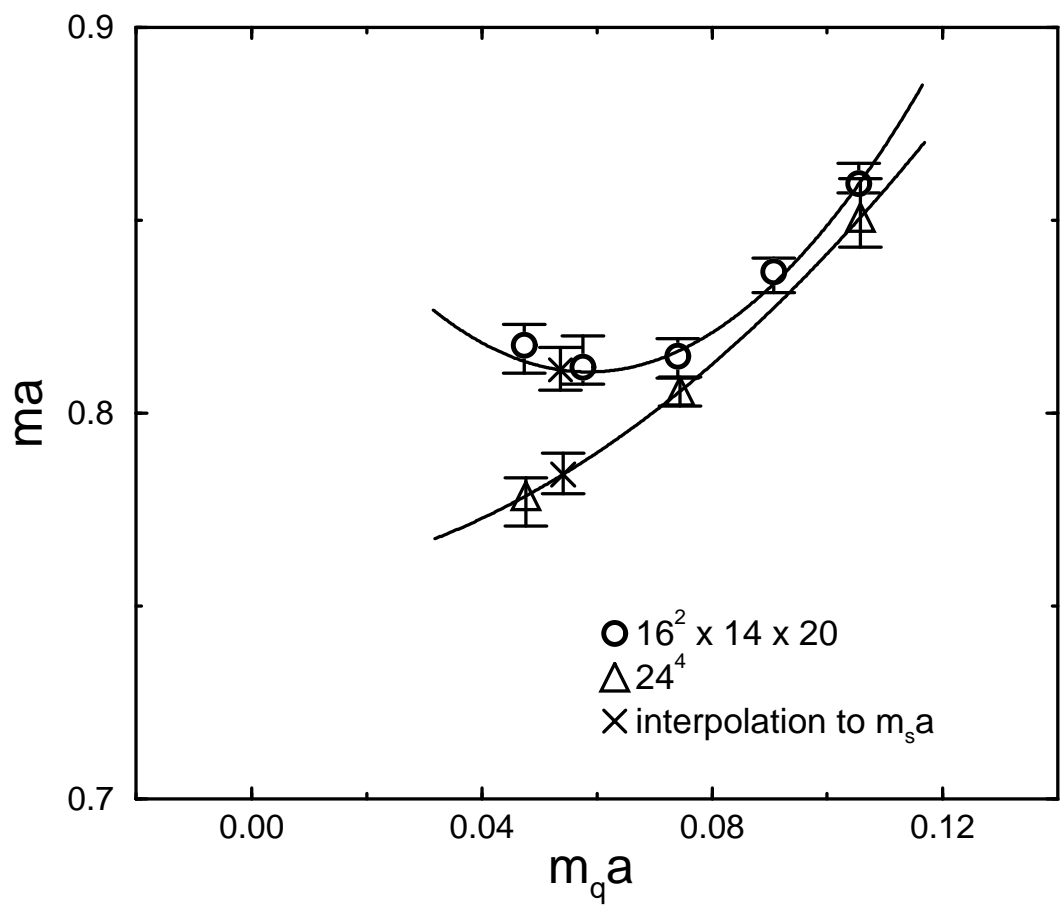}
\caption{Scalar quarkonium mass as a function of quark mass for
$\beta$ of 5.93.}
\label{fig:mass}
\end{figure}

For the two lattice with $\beta$ of 5.93, Figure~\ref{fig:mass} shows
the scalar quarkonium mass as a function of quark mass $m_q a$, defined
to be $(2 \kappa)^{-1} -(2 \kappa_c)^{-1}$. Here $\kappa$ is the hopping
constant and $\kappa_c$ is the critical hopping constant at which the
pseudoscalar mass $m_p$ goes to zero. We determined $\kappa_c$ from a
fit of $m_p^2$ to a quadratic function of $1/\kappa$.
The solid lines in
Figure~\ref{fig:mass} are quadratic fits to the scalar mass as a
function of quark mass which we used to interpolate to the strange quark mass.
The strange quark mass we chose to be the value which yields, in
physical units, a pseudoscalar mass squared of $2 m_K^2 - m_{\pi}^2$,
where $m_K$ and $m_{\pi}$ are the observed neutral kaon and pion masses,
respectively.  As shown by the figure, for the lattice $16^2 \times 14
\times 20$ with $L$ of 1.54(4) fm the scalar mass as a function of
quark mass flattens out as quark mass is lowered toward the strange
quark mass and then appears to begin to rise as the quark mass is
decreased still further.  This feature is absent from the data at
$\beta$ of 5.93 for the lattice $24^4$ with $L$ of 2.31(6) fm and is
thus a finite-volume artifact. It is present in the data at
$\beta$ of 5.70 with $L$ of 1.68(5) fm, at $\beta$ of 6.17 with $L$
of 1.74(5) fm, and at $\beta$ of 6.40 with $L$ of 1.66(5) fm, but
absent in the data at $\beta$ of 5.70 with $L$ of 2.24(7) fm.

The pseudoscalar mass squared $m_p^2$, for all values of lattice
spacing, we found to be nearly a linear function of $1/\kappa$ and
nearly independent of lattice period. The difference in $m_p a$
between the two lattice at $\beta$ of 5.70 and between the two lattice
at $\beta$ of 5.93 was in all cases less than 0.5\%.

\begin{figure}
\epsfxsize=\textwidth
\epsfbox{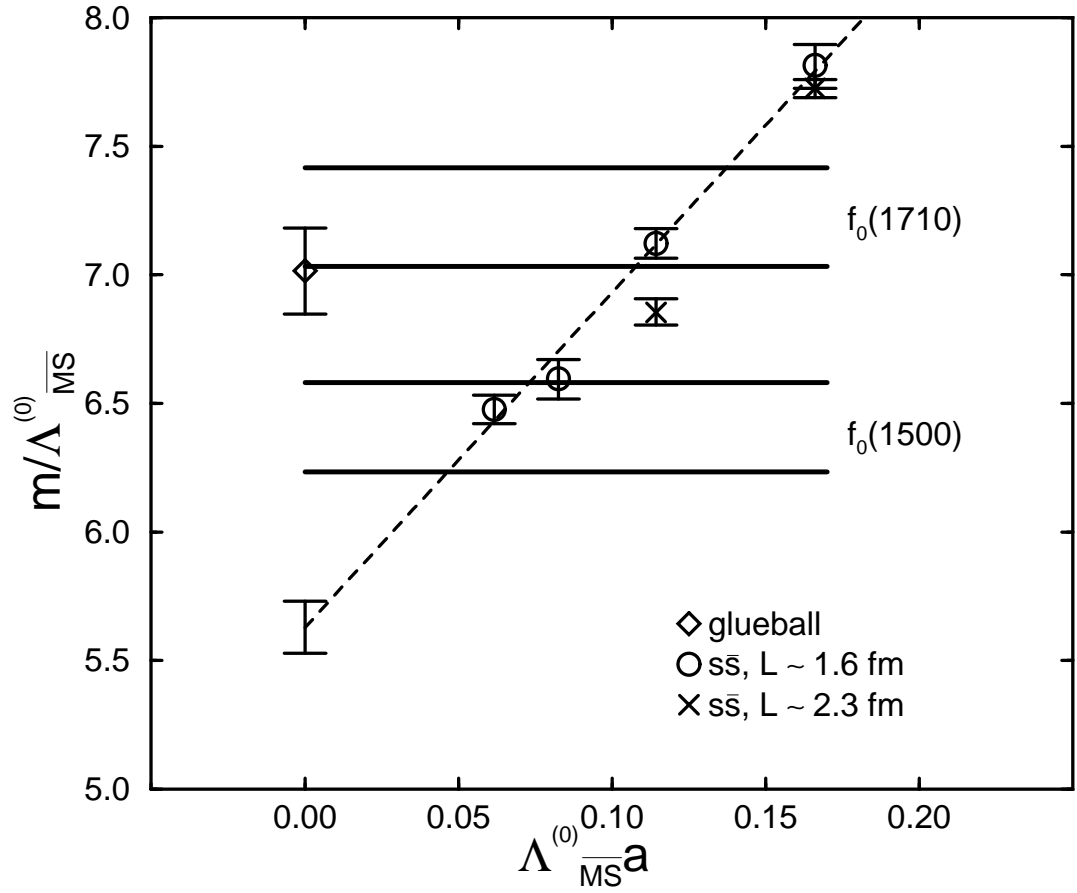}
\caption{
Lattice spacing dependence and continuum limit of the scalar
$s\overline{s}$ mass, continuum limit of the scalar glueball mass, and one
sigma upper and lower bounds on observed masses.}
\label{fig:masscont}
\end{figure}

For $L$ near 1.6 fm, Figure~\ref{fig:masscont} shows the $s\overline{s}$
scalar mass in units of $\Lambda^{(0)}_{\overline{MS}}$ as a function of
lattice spacing in units of $1/\Lambda^{(0)}_{\overline{MS}}$.  A linear
extrapolation of the mass to zero lattice spacing gives 1322(42) MeV,
far below our valence approximation infinite volume continuum glueball
mass of 1648(58) MeV.  For the ratio of the $s\overline{s}$ mass to the
infinite volume continuum limit of the scalar glueball mass we obtain
0.802(24).  Figure~\ref{fig:masscont} shows also values of the
$s\overline{s}$ scalar mass at $\beta$ of 5.70 and 5.93 with $L$ of
2.24(7) and 2.31(6) fm, respectively.  The $s\overline{s}$ mass with $L$
near 2.3 fm lies below the 1.6 fm result for both values of lattice
spacing. Thus the infinite volume continuum $s\overline{s}$ mass should
lie below 1322(42) MeV.  We believe our data rule out the interpretation
of $f_0(1500)$ as mainly composed of the lightest scalar glueball with
$f_0(1710)$ consisting mainly of $s\overline{s}$ scalar quarkonium. For
comparison with our data, Figure~\ref{fig:masscont} shows the valence
approximation value for the infinite volume continuum limit of the
scalar glueball mass and the observed value of the mass of $f_0(1500)$
and of the mass of $f_0(1710)$ The uncertainties shown in the observed
masses in units of $\Lambda^{(0)}_{\overline{MS}}$ arise mainly from the
uncertainty in $\Lambda^{(0)}_{\overline{MS}}$.

\begin{figure}
\epsfxsize=\textwidth
\epsfbox{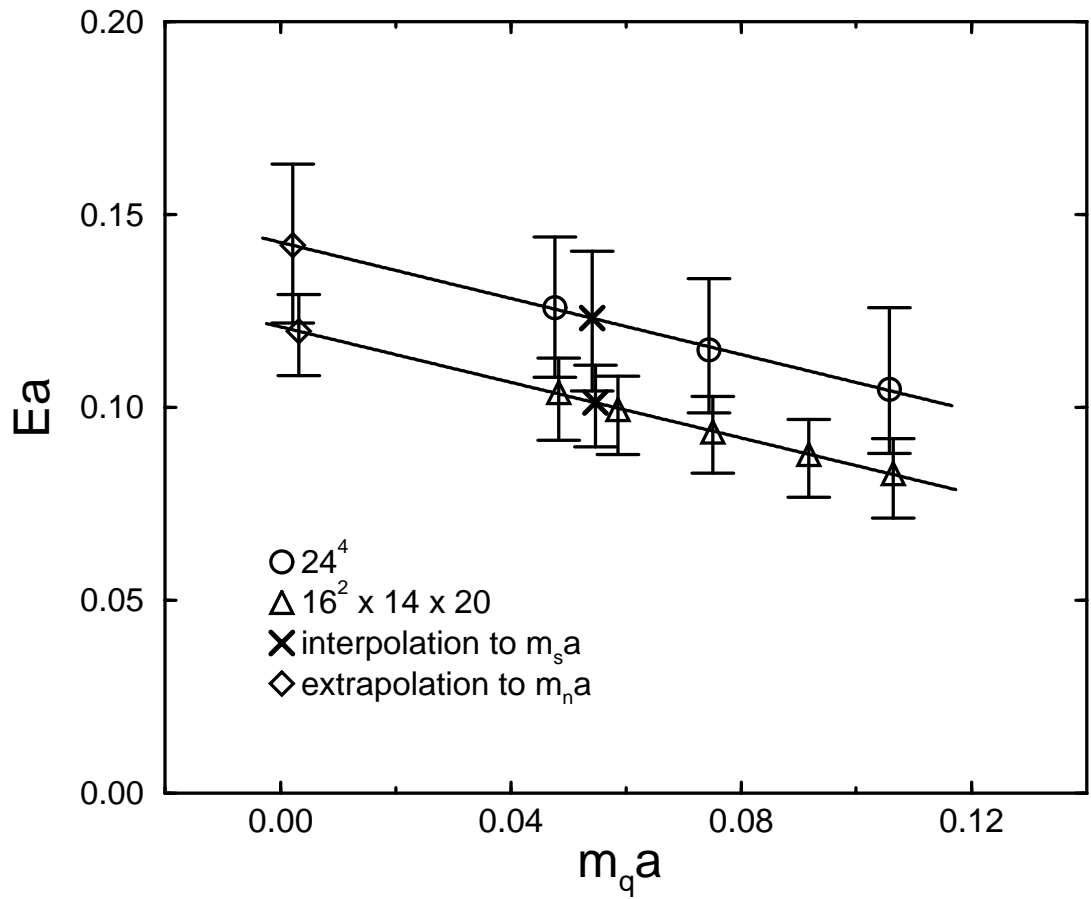}
\caption{Glueball-quarkonium mixing energy as a function of
quark mass for $\beta$ of 5.93.}
\label{fig:mix}
\end{figure}

Figure~\ref{fig:mix} shows the quarkonium-glueball mixing energy as a
function of quark mass for the two different lattices with $\beta$ of
5.93. For neither lattice does there appear to be any sign of the
anomalous quark mass dependence found in Figure~\ref{fig:mass}.  The
mixing energies at different quark masses turn out to be highly
correlated and depend quite linearly on quark mass.  For $\beta$ of
5.70, 6.17 and 6.40 the mixing energy behaves similarly and, in
particular, also depends quite linearly on quark mass.  Thus it appears
that the mixing energy can be extrapolated reliably down to the normal
quark mass $m_n$, defined to be the quark mass at which
$m_p$ becomes $m_{\pi}$.  At $\beta$ of 5.70, the mixing energy ratio
$E(m_n)/E(m_s)$ is 1.222(34) for $L$ of 1.68(5) fm and
1.194(45) for $L$ of 2.24(7) fm.  For the data at $\beta$ of 5.93, this ratio
is 1.183(32) for $L$ of 1.54(4) fm and 1.153(56) for $L$ of 2.31(6)
fm.  Thus the ratio has at most rather small volume dependence and
seems already to be near its infinite volume limit with $L$ around 1.6
fm.

\begin{figure}
\epsfxsize=\textwidth
\epsfbox{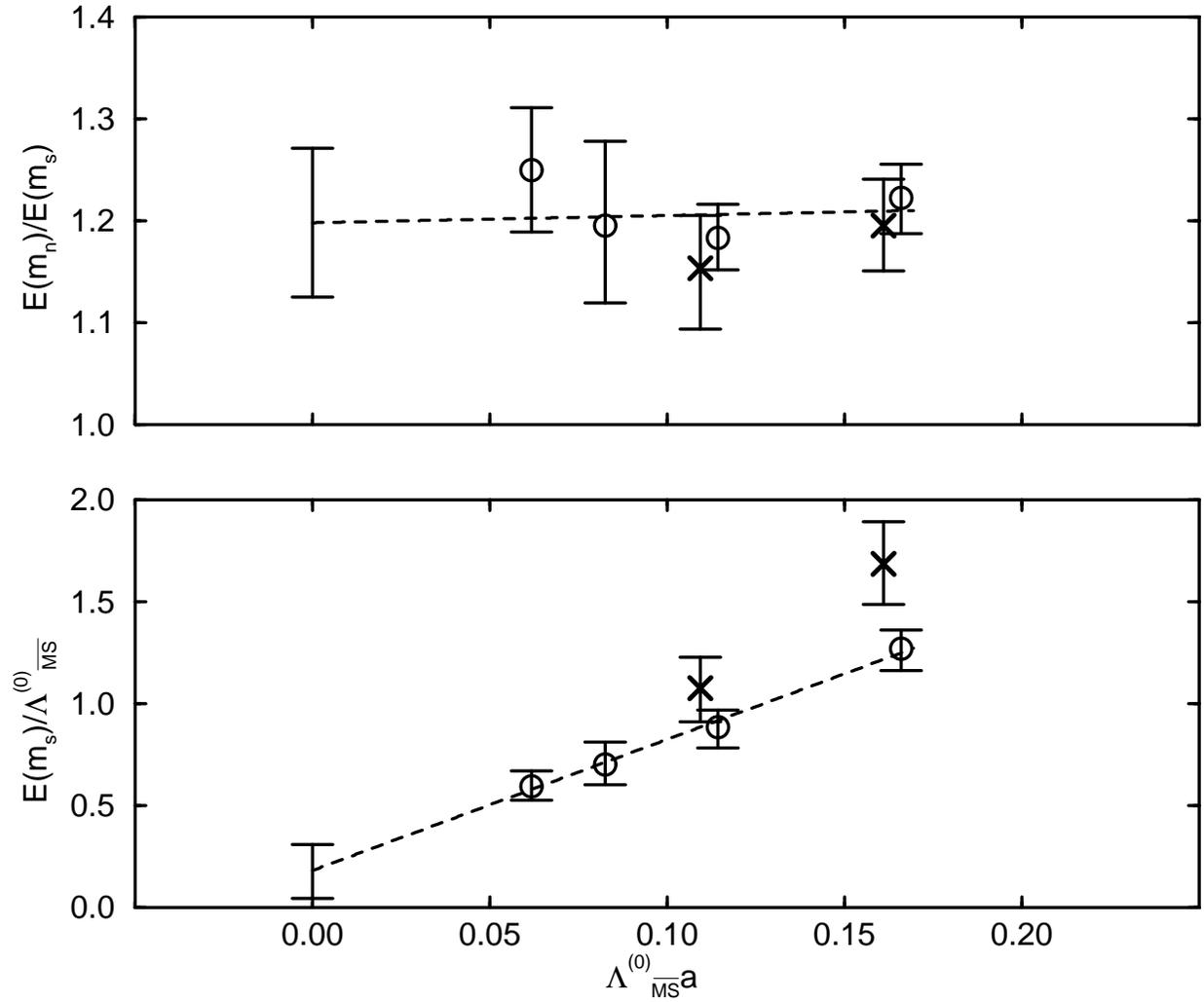}
\caption{Lattice spacing dependence and continuum limit of the
glueball-quarkonium mixing energy $E(m_s)$ and of the ratio
$E(m_n)/E(m_s)$.}
\label{fig:mixcont}
\end{figure}

Figure~\ref{fig:mixcont} shows a linear extrapolation to zero lattice
spacing of quarkonium-glueball mixing energy at the strange quark mass
$E(m_s)$ and of the ratio $E(m_n)/E(m_s)$. The limiting value
of $E(m_s)$ is 43(31) MeV and of $E(m_n)/E(m_s)$ is 1.198(72).

We now combine our infinite volume continuum value for $E(m_n)/E(m_s)$
with a simplified treatment of the mixing among valence approximation
glueball and quarkonium states which arises in full QCD from
quark-antiquark annihilation.  The simplification we introduce is to
permit mixing only between the lightest scalar glueball and the lowest
lying discrete quarkonium states.  We ignore mixing between the lightest
glueball and excited quarkonium states or multiquark continuum states,
and we ignore mixing between the lightest quarkonium states and excited
glueball states or continuum states containing both quarks and
glueballs.

Excited quarkonium and glueball states and states containing both quarks
and glueballs are expected to be high enough in mass that their effect
on the lowest lying states will be much smaller than the effect of
mixing of the lowest lying states with each other. On the other hand,
the shift in the glueball mass arising from mixing with discrete states
is, according to the systematic version of the valence approximation
described in Ref.~\cite{Lee98}, mainly a one-quark-loop correction to
the valence approximation while mixing with continuum states is mainly a
multiquark loop correction. For low energy QCD properties it is expected
multiquark loop corrections will be significantly smaller than single
loop corrections.  

The structure of the Hamiltonian coupling together the scalar glueball,
the scalar $s\overline{s}$ and the scalar $n\overline{n}$ isosinglet
becomes
\begin{displaymath}
\left| 
\begin{array}{ccc}
m_g &  E(m_s) & \sqrt{2} r E(m_s) \\
E(m_s) & m_{s\overline{s}} & 0 \\
\sqrt{2} r E(m_s) & 0 & m_{n\overline{n}}. 
\end{array} 
\right| 
\end{displaymath} 
Here $r$ is the ratio $E(m_n)/E(m_s)$ which we found to be
1.198(72), and $m_g$, $m_{s\overline{s}}$ and $m_{n\overline{n}}$
are, respectively, the glueball mass, the $s\overline{s}$ quarkonium
mass and the $n\overline{n}$ quarkonium mass before mixing.

The three unmixed mass parameters we will take as unknowns. We will also
take $E(m_s)$ as an unknown since the fractional error bar on our
measured value is large. These four unknowns can now be determined from
four observed masses. To leading order in the valence approximation,
with valence quark-antiquark annihilation turned off, corresponding isotriplet and
isosinglet states composed of $u$ and $d$ quarks will be degenerate. For
the scalar meson multiplet, the isotriplet $(u\overline{u} -
d\overline{d})/\sqrt{2}$ state has a mass reported by the Crystal Barrel
collaboration to be 1470(25) MeV~\cite{Amsler1}.  Thus we take
$m_{n\overline{n}}$ to be 1470(25) MeV.  In
addition, the Crystal Barrel collaboration finds an isosinglet mass of
1390(30) MeV~\cite{Amsler1} from one recent analysis and 
1380(40) MeV~\cite{Abele} from another.
Mark III finds 1430(40) MeV~\cite{MarkIII}.
We take the mass of the physical mixed state with largest contribution
coming from $n\overline{n}$ to be 
1404(24) MeV, the weighted average of 1390(30) MeV and 1430(40) MeV.
The mass of the physical mixed states
with the largest contributions from $s\overline{s}$ we take as the mass
of $f_0(1500)$, for which the Particle Data Group's averaged value is
1505(9) MeV.  The mass of the physical mixed state with the
largest contributions from the glueball we take as the Particle
Data group's averaged mass of $f_0(1710)$, 1697(4) MeV.

Adjusting the parameters in the matrix to give the physical eigenvalues
we just specified, $m_g$ becomes 1622(29) MeV, $m_{s\overline{s}}$
becomes 1514(11) MeV, and $E(m_s)$ becomes 64(13) MeV, with error bars
including the uncertainties in the four input physical masses. The
unmixed $m_g$ is in good agreement with world average valence
approximation glueball mass 1632(49) MeV, and $E(m_s)$ is consistent
with our measured value of 43(31) MeV.

For the three physical eigenvectors we obtain
\begin{eqnarray}
\lefteqn{| f_0(1710) >  = } \nonumber \\
& & 0.859(54) | g >  + 0.302(52) | s\overline{s}>  
+  0.413(87) |n\overline{n}>, \nonumber  \\  
\lefteqn{ | f_0(1500) >  = } \nonumber \\ 
& & -0.128(52) | g >  + 0.908(37) | s\overline{s}>  
-  0.399(113) | n\overline{n}>, \nonumber  \\ 
\lefteqn{| f_0(1390) >  = } \nonumber \\
& &  -0.495(118) | g >  + 0.290(91)  | s\overline{s}>  
+  0.819(89) | n\overline{n}>. \nonumber 
\end{eqnarray}
The mixed $f_0(1710)$ has a glueball content of 73.8(9.5)\%, the mixed
$f_0(1500)$ has a glueball content of 1.6(1.4)\% and the mixed
$f_0(1390)$ has a glueball content of 24.5(10.7)\%.  Since, as well known,
the partial width $\Gamma(J/\Psi \rightarrow \gamma + h)$ is a
measure of the size of the gluon component in the wave function of
hadron $h$, our results imply that $\Gamma(J/\Psi \rightarrow
\gamma + f_0(1710))$ should be significantly larger than $\Gamma(J/\Psi
\rightarrow \gamma + f_0(1390))$ and $\Gamma(J/\Psi \rightarrow \gamma +
f_0(1390))$ should be significantly larger than $\Gamma(J/\Psi
\rightarrow \gamma + f_0(1500))$. These predictions are supported by a
recent reanalysis of Mark III data~\cite{MarkIII}.
In addition, in the state vector for $f_0(1500)$, the relative negative sign
between the $s\overline{s}$ and $n\overline{n}$ components will lead, by
interference, to a suppression of the partial width for this state to
decay to $K\overline{K}$. Assuming SU(3) flavor symmetry for the two
pseudoscalar decay coupling of the scalar quarkonium states, the total
$K\overline{K}$ rate for $f_0(1500)$ is suppressed by a factor of
0.39(16) in comparison to the $K\overline{K}$ rate for an unmixed
$s\overline{s}$ state. This suppression is consistent, within
uncertainties with the experimentally observed suppression.

\end{document}